\documentclass[aip,jcp,amsmath,amssymb,reprint,superscriptaddress,floatfix,longbibliography]{revtex4-2}

\usepackage{graphicx}
\usepackage{physics}

\usepackage{dcolumn}
\usepackage{bm}
\usepackage[colorlinks=true,linkcolor=blue,citecolor=blue,urlcolor=blue]{hyperref}
\usepackage{amsmath,graphicx}
\usepackage{color}

\usepackage[dvipsnames]{xcolor}
\usepackage[normalem]{ulem}

\newcommand{\new}[1]{#1}

\newcommand{\ilm}{Universit{\'e} Claude Bernard Lyon 1, CNRS, Institut Lumière Matière, UMR5306, F69100 Villeurbanne, France}
\newcommand{\osaka}{Department of Mechanical Engineering, Osaka University, 2-1 Yamadaoka, Suita 565-0871, Japan}
\newcommand{\watus}{Water Frontier Research Center (WaTUS), Research Institute for Science \& Technology, Tokyo University of Science,
1-3 Kagurazaka, Shinjuku-ku, Tokyo 162-8601, Japan}

\renewcommand{\selectlanguage}[1]{}
\begin{document}

\title{Equilibrium and Non-Equilibrium Molecular Dynamics Simulation of Thermo-Osmosis: Enhanced Effects on Polarized Graphene Surfaces}

\author{Mehdi Ouadfel}
\email{mehdi.ouadfel@univ-lyon1.fr}
\affiliation{\ilm}
\author{Samy Merabia}
\email{samy.merabia@univ-lyon1.fr}
\affiliation{\ilm}
\author{Yasutaka Yamaguchi}
\email{yamaguchi@mech.eng.osaka-u.ac.jp}
\affiliation{\osaka}
\affiliation{\watus}
\author{Laurent Joly}
\email{laurent.joly@univ-lyon1.fr}
\affiliation{\ilm}

\date{\today}

\begin{abstract}
Thermo-osmotic flows, generated by applying a thermal gradient along a liquid-solid interface, could be harnessed to convert waste heat into electricity. While this phenomenon has been known for almost a century, there is a crucial need to gain a better understanding of the molecular origins of thermo-osmosis. In this paper, we start by detailing the multiple contributions to thermo-osmosis. We then showcase three approaches to compute the thermo-osmotic coefficient using molecular dynamics; a first method based on the computation of the interfacial enthalpy excess and Derjaguin's theoretical framework, a second approach based on the computation of the interfacial entropy excess using the so-called dry-surface method, and a novel non-equilibrium method to compute the thermo-osmotic coefficient in a periodic channel. We show that the three methods align with each other, in particular for smooth surfaces. In addition, for a polarized graphene-water interface, we observe large variations of thermo-osmotic responses, and multiple changes in flow direction with increasing surface charge. Overall, this study showcases the versatility of osmotic flows and calls for experimental investigation of thermo-osmotic behavior in the vicinity of charged surfaces.

\end{abstract}

\maketitle

\section{Introduction}

Nanofluidic systems offer great promises for energy harvesting \cite{schoch_transport_2008,bocquet_nanofluidics_2010,kavokine_fluids_2021}. At the nanoscale, osmotic flows, which are flows generated by a thermodynamic gradient along solid-liquid interfaces, can be used to generate electricity. Indeed, if one considers an aqueous electrolyte near a charged surface, the ions in the liquid will move under the influence of the electric field generated by a surface to form the electrical double layer (EDL), a layer of electrically charged liquid \cite{grahame_electrical_1947,IsraelachviliBook,hartkamp_measuring_2018}. The advection of the EDL by osmotic flows creates an electric current. For example, diffusio-osmosis, a phenomenon wherein the flow is generated by a salinity gradient, creates a so-called diffusio-osmotic current \cite{siria_giant_2013,fair_reverse_1971,mouterde_interfacial_2018}.

One can also generate an osmotic flow by applying a thermal gradient along an interface, which is referred to as thermo-osmosis \cite{barragan_thermo-osmosis_2017,wurger_thermal_2010,herrero2022chapter}. Derjaguin and colleagues developed a theoretical framework for thermo-osmosis, which relates the thermo-osmotic flow to the interfacial enthalpy excess \cite{derjaguin_surface_1987,Derjaguin1941,Anderson1989}. This approach, based on linear irreversible thermodynamics and continuum hydrodynamics, faces several issues. The enthalpy excess, originating from the interfacial layer (interacting with the wall), typically spans a range on the order of 1\,nm. In this context, the validity of the equation of continuum hydrodynamics becomes questionable \cite{travis_departure_1997}. Moreover, Derjaguin's approach assumes a constant viscosity of the fluid near the surface. While this assumption is reasonable for hydrophobic surfaces \cite{sendner_interfacial_2009,bonthuis_beyond_2013,li_structured_2007}, it is generally not true \cite{ganti_molecular_2017,li_structured_2007,bonthuis_beyond_2013,leng_fluidity_2005}. Moreover, the local enthalpy excess is calculated using the pressure tensor, which is not uniquely defined when considering heterogeneous liquids at interfaces \cite{shi_perspective_2023}.

Recently, numerous theoretical works brought together the macroscopic description of thermo-osmosis with a molecular description of the interfacial region \cite{ganti_molecular_2017,ganti_hamiltonian_2018,anzini_fluid_2022,herrero_fast_2022}. Using Onsager's reciprocity relation, which links the flow generated by a thermal gradient to the heat flux due to a pressure gradient, these studies have been able to study thermo-osmosis \textit{via} the 'mechano-caloric' route \cite{fu_what_2017,herrero_fast_2022,chen_thermo-osmosis_2021,chen_thermo-osmosis_2023}. A good agreement was found between this approach and Derjaguin's approach for model systems \cite{fu_what_2017,proesmans_comparing_2019}.
However, whether Derjaguin's framework can be applied at the molecular scale to predict thermo-osmotic transport remains under debate.

In this paper, we investigate the thermo-osmotic coefficient of realistic solid-liquid interfaces, \textit{i.e.}, water on graphitic surfaces, using molecular dynamics (MD) simulations. We begin by detailing the different contributions to thermo-osmosis. Subsequently, we calculate the thermo-osmotic coefficient using three different methods. First, we employ a previously discussed approach based on calculating the enthalpy excess at equilibrium and Derjaguin's theoretical framework \cite{ouadfel_complex_2023}. Second, we introduce an alternative approach that involves computing the solid-liquid entropy excess at equilibrium using the so-called dry-surface method (DSM). This method enables the calculation of the solid-liquid work of adhesion, which is linked to the entropy excess, through thermodynamic integration along the solid-liquid interaction parameter. \cite{surblys_molecular_2018,leroy_dry-surface_2015,ardham_solid-liquid_2015}. Finally, we have developed a non-equilibrium MD-based method to compute the thermo-osmotic response in a periodic channel, in which we directly apply a thermal gradient and compute the thermo-osmotic velocity. We find a remarkable agreement between the different approaches, confirming that it is possible to apply Derjaguin theory at the molecular scale. In addition, we observe large variations of thermo-osmotic responses on graphitic surfaces, large responses in polarized graphene, and multiple changes in flow direction with increasing surface charge.

\section{Theory}\label{sec:theory}
\begin{figure}
    \centering
    \includegraphics[width=0.45\textwidth]{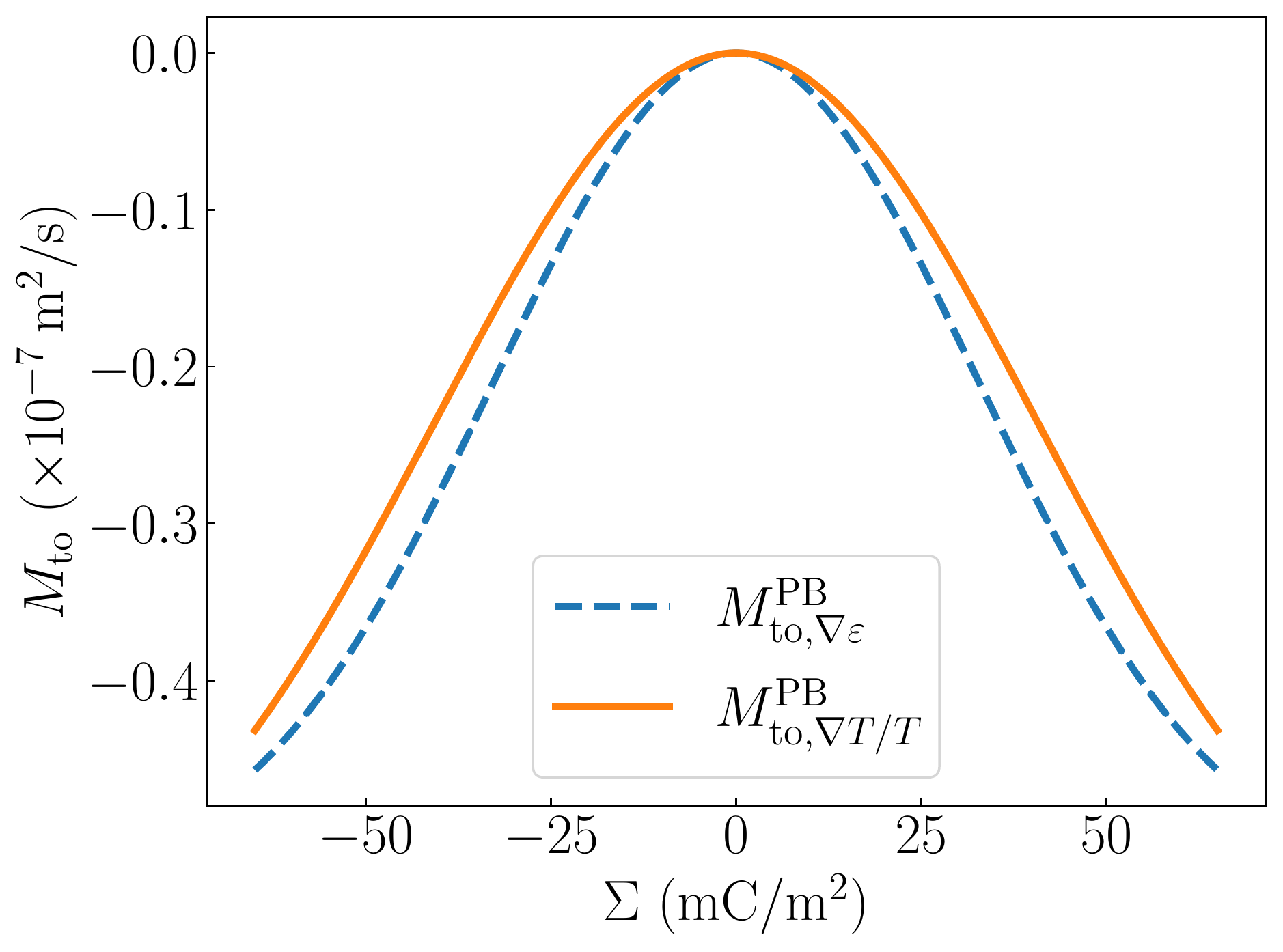}
    \caption{Comparison of the contributions of the thermal and permittivity gradients to thermo-osmosis in the Poisson-Boltzmann theory for a polarized graphene surface\new{, calculated for a salt concentration of 0.3\,M and a slip length of 50\,nm, representative of the graphene systems considered in the MD simulations}.}
    \label{fig:mto_pb_comparison}
\end{figure}

In this section, we will showcase the different contributions to thermo-osmosis. \new{Following the work of Derjaguin and collaborators \cite{derjaguin_surface_1987,Derjaguin1941,Anderson1989},} taking into account solid-liquid slip, one can solve the Stokes equation and obtain the velocity of the fluid far from the surface, \textit{i.e.} the osmotic velocity \cite{herrero_fast_2022,fu_what_2017,herrero2022chapter}:
\begin{equation}\label{eq:osmotic_velocity}
    v_\mathrm{osm} = \frac{1}{\eta}\int_0^\infty (z+b)f_x(z)\dd z,
\end{equation}
in which $z=0$ is the position of the surface, $z=\infty$ corresponds to the middle of the channel, $\eta$ is the viscosity, considered homogeneous, $b$ is the slip length \cite{bocquet_flow_2007}, which quantifies solid-liquid slip, and $f_x$ is the force density exerted on the fluid in the $x$ direction. When a thermal gradient is applied along the interface, this force density is written \cite{wurger_thermal_2010}
\begin{equation}
    f_x(z)=-\delta h(z)\frac{\nabla T}{T} - \frac{E(z)^2}{2}\nabla\varepsilon.
\end{equation}
The first term can be derived from local thermodynamic equilibrium \cite{bregulla_thermo-osmotic_2016,ganti_molecular_2017}, it is the force density generated by a temperature gradient, wherein $\delta h(z)$ is the enthalpy excess density and $\nabla T$ is the temperature gradient.
The second term, which we could refer to as temperature-dependent permittivity-osmosis, arises from a gradient of dielectric permittivity $\varepsilon$. Note that this term, at constant pressure, depends solely on the variation in permittivity induced by temperature
\cite{catenaccio_temperature_2003}. Here, $E(z)$ is the electric field, which is approximately equal to its transverse component $E_z(z)$ close to an electrically charged surface. 
Considering only the ion contribution to the enthalpy excess, using the Poisson-Boltzmann (PB) framework (here for a symmetric $Z$:$Z$ salt), 
the force density becomes:
\begin{multline}
    f_x(z)=-\left[-\varepsilon V(z)\dv[2]{V}{z}+\frac{\varepsilon}{2}\left(\dv{V}{z}\right)^2\right]\frac{\nabla T}{T}\\
    -\frac{1}{2}\left(\dv{V}{z}\right)^2\nabla \varepsilon,
\end{multline}
where $V(z)$, the electrostatic potential, is given by the PB theory \cite{herrero_poisson-boltzmann_2022,herrero_fast_2022,blossey_poisson-boltzmann_2023}. One can then integrate Eq.~(\ref{eq:osmotic_velocity}) to obtain the thermo-osmotic velocity
\begin{equation}
    v_\mathrm{to}^{\mathrm{PB}}=v_{\mathrm{to},\nabla T}^{\mathrm{PB}}+v_{\mathrm{to},\nabla\varepsilon}^{\mathrm{PB}},
\end{equation}
with
\begin{multline}
    v_{\mathrm{to},\nabla T}^{\mathrm{PB}}=-\frac{\nabla T/T}{2\pi\ell_{\mathrm{B}}\eta\beta}\biggl\{-3\ln\biggl(1-\gamma^{2}\biggr)-\mathrm{asinh}^{2}(x)\\
    +\frac{b}{\lambda_{\mathrm{D}}}\biggl[3x|\gamma|-2x\,\mathrm{asinh}(x)\biggr]\biggr\},
\end{multline}
and
\begin{equation}\label{eq:vto_eps}
    v_{\mathrm{to},\nabla\varepsilon}^{\mathrm{PB}}=-\frac{\nabla\varepsilon/\varepsilon}{2\pi \ell_\mathrm{B}\eta\beta}\left[-\ln(1-\gamma^2)+\frac{b}{\lambda_\mathrm{D}}x|\gamma|\right],
\end{equation}
where $\ell_\mathrm{B}=\beta q^2/(4\pi\varepsilon)$ is the Bjerrum length, with $\beta=1/(k_\mathrm{B}T)$ and $q=Ze$ the absolute ionic charge, $\lambda_\mathrm{D}=1/(\sqrt{8\pi\ell_\mathrm{B}n_0})$ is the Debye length, in which $n_0$ is the bulk ion concentration, and $x=\lambda_\mathrm{D}/\ell_\mathrm{GC}$, where $\ell_\mathrm{GC}=q/(2\pi\ell_\mathrm{B}|\Sigma|)$ is the Gouy-Chapman length, with $\Sigma$ the surface charge density. Finally
\begin{equation}
    \gamma=\frac{\mathrm{sgn}(\Sigma)}{x}\left[-1+\sqrt{1+x^2}\right].
\end{equation}
The permittivity gradient can be expressed as a function of the thermal gradient \cite{wurger_thermal_2010}:
\begin{equation}
    \frac{\nabla\varepsilon}{\varepsilon}=-\tau\frac{\nabla T}{T},
\end{equation}
with $\tau=1.4$ for water at room temperature. Using this relation, one can compute the two contributions to thermo-osmosis, which we quantify by using the thermo-osmotic coefficient:
\begin{equation}\label{eq:mto_nemd}
    M_\mathrm{to}=-\frac{v_\mathrm{to}}{\nabla T/T}.
\end{equation}
Consequently, $M_{\mathrm{to},\nabla T}^{\mathrm{PB}}=-v_{\mathrm{to},\nabla T}^{\mathrm{PB}}/(\nabla T/T)$ and $M_{\mathrm{to},\nabla\varepsilon}^{\mathrm{PB}}=-v_{\mathrm{to},\nabla\varepsilon}^{\mathrm{PB}}/(\nabla T/T)$. Figure~\ref{fig:mto_pb_comparison} shows that $M_{\mathrm{to},\nabla T}^{\mathrm{PB}}$ and $M_{\mathrm{to},\nabla\varepsilon}^{\mathrm{PB}}$ are comparable in the PB theory. However, it has been shown that the PB contribution to enthalpy excess can be negligible compared to that of water near some surfaces \cite{herrero_fast_2022,ouadfel_complex_2023}. In this case, the effect of the permittivity gradient is also negligible compared to that of the enthalpy excess term of water.

In the general case, for a mixture of particles, the enthalpy excess density is defined as \cite{ganti_molecular_2017}:
\begin{equation}\label{eq:enthalpy_excess_density}
    \delta h(z)=\sum_in_i(z)[h_i(z)-h_i^\mathrm{bulk}],
\end{equation}
with $n_i$ and $h_i$ respectively the number density and the enthalpy per particle of the species $i$, and the superscript bulk refers to a quantity far from the surface, where it is homogeneous. The enthalpy excess is given by:
\begin{equation}
    \Delta H=\int_0^\infty\delta h(z)\dd z.
\end{equation}
Note that this definition differs from the classical definition of an excess quantity in the surface thermodynamics framework \cite{rowlinson_molecular_2002}. 
One can then rewrite $M_{\mathrm{to},\nabla T}$ as:
\begin{align}
    M_{\mathrm{to},\nabla T}&=-\frac{v_{\mathrm{to},\nabla T}}{\nabla T/T}\\
    &=\frac{1}{\eta}\int_0^\infty(z+b)\delta h(z)\dd z\\
    &=\frac{\Delta H}{\eta}(\lambda_\mathrm{h}+b),\label{eq:mto_derjaguin}
\end{align}
where
\begin{equation}
    \lambda_\mathrm{h}=\frac{1}{\Delta H}\int_0^\infty z\delta h(z)\dd z
\end{equation}
denotes the extent of the layer where the liquid interacts with the wall, and is typically around 7\,\AA{} \cite{ouadfel_complex_2023}. Thus, $M_\mathrm{to}$ can be estimated by computing $\Delta H$ at equilibrium, $\eta$, and $b$.

Another mean of computing the thermo-osmotic coefficient is to calculate the entropy excess generated by the interactions between the liquid and the surface. Indeed one can relate the enthalpy excess to the entropy excess. 
Introducing $\mu_i$, the chemical potential of the species $i$, and $s_i$, the entropy per particle of the species $i$, one can use the relation $\mu_i=h_i-Ts_i$ and the fact that $\mu_i$ and $T$ are homogeneous along $z$ to show that \cite{ganti_molecular_2017}: 
\begin{equation}
    \delta h(z)=T\delta s(z),
\end{equation}
with the entropy excess density defined as:
\begin{equation}
    \delta s(z)=\sum_in_i(z)[s_i(z)-s_i^\mathrm{bulk}].
\end{equation}
The entropy excess thus writes:
\begin{equation}
    \Delta S=\int_0^\infty\delta s(z)\dd z,
\end{equation}
and it follows that
\begin{equation}
    \Delta H=T\Delta S.
\end{equation}
Therefore, one can write:
\begin{equation}\label{eq:mto_entropy}
    M_\mathrm{to}=\frac{T\Delta S}{\eta}(b+\lambda_\mathrm{h}).
\end{equation}

One can relate this result to the flow generated by a Marangoni stress. Indeed, the Marangoni stress can be written as a function of the enthalpy excess (and thus the entropy excess) \cite{ganti_hamiltonian_2018}:
\begin{equation}
    \nabla\gamma=\left(\frac{\partial\gamma}{\partial T}\right)_P\nabla T=-\Delta H\frac{\nabla T}{T}=-\Delta S\nabla T,
\end{equation}
and the osmotic velocity becomes
\begin{equation}
    v_\mathrm{to}=\frac{\nabla\gamma}{\eta}(\lambda_\mathrm{h}+b).
\end{equation}
This result illustrates an equivalence between Marangoni and thermo-osmotic flows. On low friction surfaces where $\lambda_\mathrm{h}\ll b$, the velocity scales as $\sim\nabla\gamma\, b/\eta$, which is reminiscent of the scaling law of the Marangoni flow for superhydrophobic surfaces \cite{gao_optical_2021}.

\section{Methods}

\subsection{Systems and force field}
We computed the thermo-osmotic coefficient of an aqueous electrolyte confined between graphitic walls using MD with the LAMMPS package \cite{LAMMPS2022}. We used sodium chloride (NaCl) as the salt, with a bulk concentration $n_0 \sim 0.3$\,M, corresponding to a Debye length $\lambda_\mathrm{D} \sim 5.7$\,\AA{}. We simulated the aqueous electrolyte using the \mbox{Madrid-2019} force field \cite{zeron_force_2019}, based on TIP4P/2005 \cite{abascal_general_2005}, a rigid non-polarizable water model, and scaled charged for the ions, $|q_\mathrm{Na}|=|q_\mathrm{Cl}|=0.85$\,e. The walls were kept frozen. \new{Unless specified,} for water-\new{carbon} interactions, we took the parameters from Ref.~\citenum{falk_molecular_2010}, $\varepsilon_\mathrm{CO}=0.114$\,kcal/mol and $\sigma_\mathrm{CO}=3.28$\,\AA{}, where C and O correspond to the carbon and oxygen atoms, respectively\new{; this corresponds to a contact angle of around 85$^\circ$, see Appendix~B}. 
To assess our theoretical predictions, we also considered graphene-like surfaces with different wetting properties, which was achieved by modifying $\varepsilon_\mathrm{CO}$\new{, see Appendix~B for details}. We applied the Lorentz-Berthelot mixing rules to set the carbon-ions interaction parameters. We truncated the LJ interactions as well as the Coulombic interactions at 10\,\AA{}. For the Coulombic interactions, we used the particle-particle particle-mesh (PPPM) solver for long-range corrections, with a relative error in forces of $10^{-4}$. We used periodic boundary conditions along the $x$ and $y$ directions.

\begin{figure}
    \centering
    \includegraphics[width=0.47\textwidth]{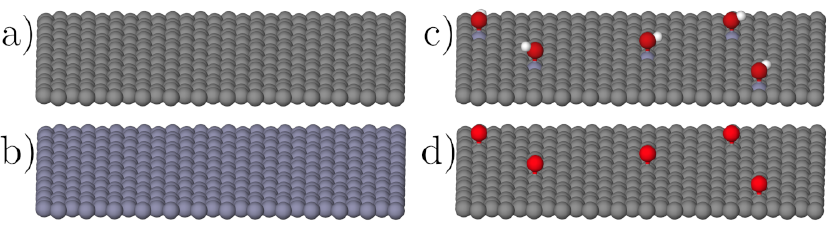}
    \caption{Different types of graphitic surfaces considered: a neutral pristine graphene (PG) surface (a), a polarized PG surface (b), a neutral graphene oxide (GO) surface (c), and an electrically charged GO surface, for which the hydrogen from the hydroxyl groups are removed to create charged groups (d). The snapshots were produced with the software Ovito \cite{stukowski_visualization_2010}.}
    \label{fig:system}
\end{figure}

We replicated the experimental structure of graphene, with an inter-atomic distance of 1.42 \AA{} \cite{yang_structure_2018}. To study the impact of surface defects on thermo-osmotic flows, we considered two types of graphene surfaces (Fig.~\ref{fig:system}): pristine graphene (PG) and graphene oxide (GO). We modeled GO by randomly distributing hydroxyl groups on the surface, with various oxidation rates $\xi=N_\mathrm{O}/N_\mathrm{C}$, where $N_\mathrm{O}$ and $N_\mathrm{C}$ are the number of hydroxyl and carbon atoms on the surface, respectively. This approach aligns with established methodologies in previous MD papers \cite{bahamon_molecular_2019,wei_breakdown_2014,giri_salt_2019}. The interaction parameters for these hydroxyl groups were based on the phenol parameters within the all-atom optimized potentials for liquid simulation (OPLS-AA) force field \cite{jorgensen_development_1996}.

The graphene sheets were either kept neutral or charged. To charge the PG surface uniformly, we assigned the same charge to all carbon atoms $q = \Sigma S/N_\mathrm{wall}$, where $S$ is the surface area of the sheet, and $N_\mathrm{wall}$ is the number of carbon atoms, resulting in a surface charge density $\Sigma$. To charge the GO, we remove the hydrogen atoms from the hydroxyl groups. In practice, this was achieved by setting the charge of the carbon, oxygen and hydrogen of the hydroxyl groups to $q_\mathrm{C_{phenol}}=0$\,e, $q_\mathrm{O_{phenol}}=-0.85$\,e and $q_\mathrm{H_{phenol}}=0$\,e, respectively, to match the rescaled charge of the ions. Counter-ions were added to the system to keep it electrically neutral.

In all simulations, we froze the bottom surface. We used the top surface as a piston during an equilibration phase that lasted 1\,ns, before fixing it at its equilibrium position to set the pressure to 1\,atm, for all simulations except the DSM (see details of DSM simulations in Section~\ref{sec:dsm}).

For the systems with a temperature gradient, we used a NVE time integration, and fixed the fluid temperature at 280 and 360\,K for the cold and hot region, respectively, \textit{via} a canonical sampling thermostat that uses global velocity rescaling with Hamiltonian dynamics \cite{bussi_canonical_2007}, applied in the $y$ and $z$ directions, following a previous thermo-osmosis study \cite{anzini_fluid_2022}, with a damping time of 200\,fs. The mean temperature was 320\,K. For other systems, we used a Nosé-Hoover thermostat to set the temperature to 320\,K with a damping time of 200\,fs. For all simulations, the timestep was 2\,fs.

\subsection{Enthalpy excess route}
To assess the predictions of Derjaguin's theory, we calculated the thermo-osmotic coefficient using Eq.~(\ref{eq:mto_derjaguin}). To this end, we computed the enthalpy excess by following the procedure described in Ref.~\citenum{ouadfel_complex_2023}, with a system composed of 1500 water molecules and 9 pairs of ions, confined between graphene walls of dimensions $L_x\sim34$\,\AA{} and $L_y\sim26$\,\AA{}, for a separation between the walls of approximately 50\,\AA{}. The enthalpy excess density is computed with Eq.~(\ref{eq:enthalpy_excess_density}), in which the enthalpy per particle is defined as \cite{ouadfel_complex_2023}:
\begin{equation}
    h_i(z)=u_i(z)+\frac{p^\parallel(z)}{n_\mathrm{tot}(z)},
\end{equation}
with $u_i$ the internal energy of species $i$, $n_\mathrm{tot}$ the total number density, and $p^\parallel=p^{xx}=p^{yy}$ the components of the virial pressure tensor parallel to the surface. Indeed, the pressure is anisotropic near the surface, and one must consider the parallel components of the pressure tensor to compute the enthalpy density  \cite{anzini_fluid_2022,ganti_molecular_2017,ganti_hamiltonian_2018}\new{; see in particular a detailed discussion in the supplementary material of Ref.~\citenum{ganti_hamiltonian_2018}}.
\new{To estimate the error on the enthalpy excess, we computed the statistical error based on independent measurements.}

We also calculated the slip length for these systems. The slip length is defined as \cite{bocquet_green-kubo_2013}:
\begin{equation}
    b = \frac{\eta}{\lambda},
\end{equation}
where $\eta$ is the viscosity of the bulk liquid, and $\lambda$ is the solid-liquid friction coefficient. One way to compute these quantities is to use Green-Kubo formulas, at equilibrium. \new{We computed the viscosity in independent simulations of a bulk liquid, using} the Green-Kubo relation \cite{alfe_first-principles_1998}:
\begin{equation}
    \eta=\frac{V}{k_\mathrm{B}T}\int_0^\infty\langle p_{\alpha\beta}(t)p_{\alpha\beta}(0)\rangle\dd t,
\end{equation}
with $V$ the volume of the simulation box, and $p_{\alpha\beta}$ the independent components of the traceless pressure tensor. Similarly, the friction coefficient is calculated at equilibrium using \cite{bocquet_green-kubo_2013}:
\begin{equation}
    \lambda=\frac{1}{Ak_\mathrm{B}T}\int_0^\infty\langle F_\alpha(t)F_\alpha(0)\rangle\dd t,
\end{equation}
where $A$ is the wall surface and $F$ is the force acting on the wall along the $\alpha$ direction, $\alpha=x,y$.

One can also compute the slip length using NEMD simulations and 
the Navier boundary condition 
\cite{Cross2018, xie_liquid-solid_2020}:
\begin{equation}
    b=\frac{v_\mathrm{s}}{\dot{\gamma}},
\end{equation}
where $\dot{\gamma}$ is the bulk shear rate and $v_\mathrm{s}$ is the slip velocity, defined as the difference between the wall velocity and the velocity of the fluid at the position of the hydrodynamic wall, given by $\dot{\gamma}h/2$, where $h$, the hydrodynamic height of the liquid, is given by \cite{herrero_shear_2019}:
\begin{equation}
    h=\frac{M}{\rho_\mathrm{bulk}A},
\end{equation}
with $M$ the total mass of the fluid and $\rho_\mathrm{bulk}$ the bulk mass density.
To compute those quantities, we moved the walls in opposite parallel directions, at a constant speed $V_x\in[10,50]$ m/s, generating a linear velocity profile far from the wall. We verified that the resulting quantities remained in the linear response regime.
\new{Green-Kubo measurements were preferred for systems with high slip lengths, while non-equilibrium simulations were used for smaller slip lengths.}

\subsection{Entropy excess route (DSM)}\label{sec:dsm}
The entropy excess can be derived from the solid-liquid entropy per unit area $s_\mathrm{sl}$, which can be computed using the DSM. It is based on the thermodynamic integration of the solid-liquid work of adhesion  $W_\mathrm{sl}$ along a coupling parameter $\kappa$, that quasi-statically weakens the solid-liquid LJ interactions to a point $\kappa_0$ where the liquid is no longer influenced by the solid. Here, the system is only composed of one graphitic surface and an aqueous electrolyte. The entropy difference is computed using the following relation \cite{surblys_molecular_2018}:
\begin{equation}\label{eq:wsl}
    W_\mathrm{sl}(\kappa)=\delta u_\mathrm{sl}(\kappa)-T\delta s_\mathrm{sl}(\kappa).
\end{equation}
Here, $\delta A(\kappa)=A(\kappa)-A(\kappa_0)$ is the difference of the quantity $A$ between its value at $\kappa$ and a reference state, defined by $\kappa_0$.

The work of adhesion can be expressed in terms of Helmholtz free energy:
\begin{align}
    W_\mathrm{sl}(\kappa)&=\left(\frac{\Delta F}{A}\right)_{N,V,T}\\
    &=-\frac{1}{A}[F(\kappa)-F(\kappa_0)]\\
    &=-\frac{1}{A}\int_{\kappa_0}^\kappa\left\langle\frac{\partial\phi(\kappa')}{\partial\kappa'}\right\rangle_{N,V,T}\dd\kappa',\label{eq:wsl_int}
\end{align}
with $\phi$ the solid-liquid potential. The surface is not charged here, thus we only have LJ solid-liquid interactions:
\begin{equation}
    \phi_{\mathrm{sl(LJ)}}(\kappa)=\kappa\sum_{i\in\mathrm{liquid}}\sum_{j\in\mathrm{wall}}4\varepsilon_{ij}\left[\left(\frac{\sigma_{ij}}{r_{ij}}\right)^{12}-\left(\frac{\sigma_{ij}}{r_{ij}}\right)^{6}\right],
\end{equation}
its derivative is given by
\begin{equation}
    \frac{\partial\phi_\mathrm{sl(LJ)}(\kappa)}{\partial\kappa}=\phi_\mathrm{sl(LJ)}(\kappa=1).
\end{equation}
In practice we assume ergodicity and we substitute ensemble averages by time averages:
\begin{equation}
    \left\langle\frac{\partial\phi(\kappa)}{\partial\kappa}\right\rangle=\overline{\frac{\partial\phi(\kappa)}{\partial\kappa}}=\frac{1}{t_\mathrm{sim}}\int_0^{t_\mathrm{sim}}\frac{\partial\phi(\kappa)}{\partial\kappa}\dd t.
\end{equation}
The integration was performed on 30 points, $\kappa\in[0,1.5]$, to cover all the wettings used in NEMD. For each point, the equilibration phase lasted 0.6\,ns and we saved the trajectory of the simulation for 1\,ns, with a period of 1\,ps. We then applied Eq.~(\ref{eq:wsl_int}) on those trajectory files, by setting $\kappa$ to 1, with the ``rerun'' command of LAMMPS.

We also computed $\delta u_\mathrm{sl}(\kappa)$ with the rerun command, by turning off all interactions parameters except for the solid-liquid ones, and computing the total energy per unit area. We thus computed $\delta s_\mathrm{sl}$ using Eq.~(\ref{eq:wsl}). Finally,
\begin{equation}\label{eq:dS_dsm}
    s_\mathrm{sl}(\kappa)=\Delta S=\delta s_\mathrm{sl}(\kappa)+s_\mathrm{sl}(\kappa_0),
\end{equation}
in which we have considered the solid-liquid entropy to be equal to the entropy excess. We discuss this hypothesis in Appendix~A. The solid-liquid entropy of the reference state $s_\mathrm{sl}(\kappa_0)$ is unknown. We choose the value of $s_\mathrm{sl}(\kappa_0)$ so that the solid-liquid entropy $s_\mathrm{sl}(\kappa)$ vanishes for $\varepsilon_\mathrm{CO}=0.105$ kcal/mol, to have a vanishing entropy excess when $\Delta H = 0$. 
\new{To estimate the error on the entropy excess, we computed the statistical error based on block averaging of the measured values during a single simulation.}

\subsection{Non-equilibrium computation of the thermo-osmotic coefficient}
\begin{figure}
    \centering
    \includegraphics[width=0.47\textwidth]{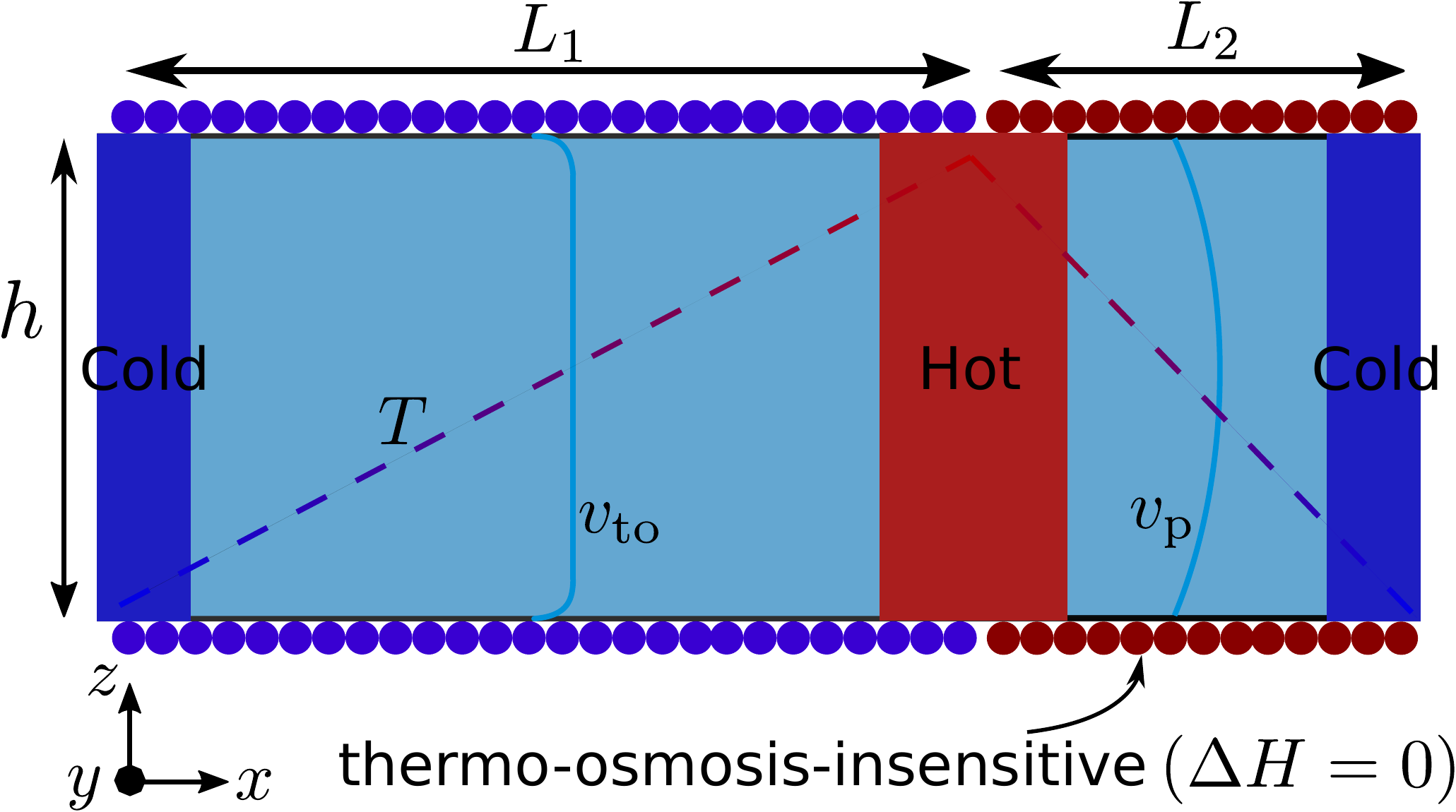}
    \caption{System used for the NEMD computation of the thermo-osmotic coefficient. A temperature gradient generates a thermo-osmotic flow in region 1, it is converted to a Poiseuille flow in region 2, because the surface in this region is thermo-osmosis-insensitive, \textit{i.e.} the enthalpy excess is null for this surface.}
    \label{fig:nemd_system}
\end{figure}
\begin{figure}
    \centering
    \includegraphics[width=0.45\textwidth]{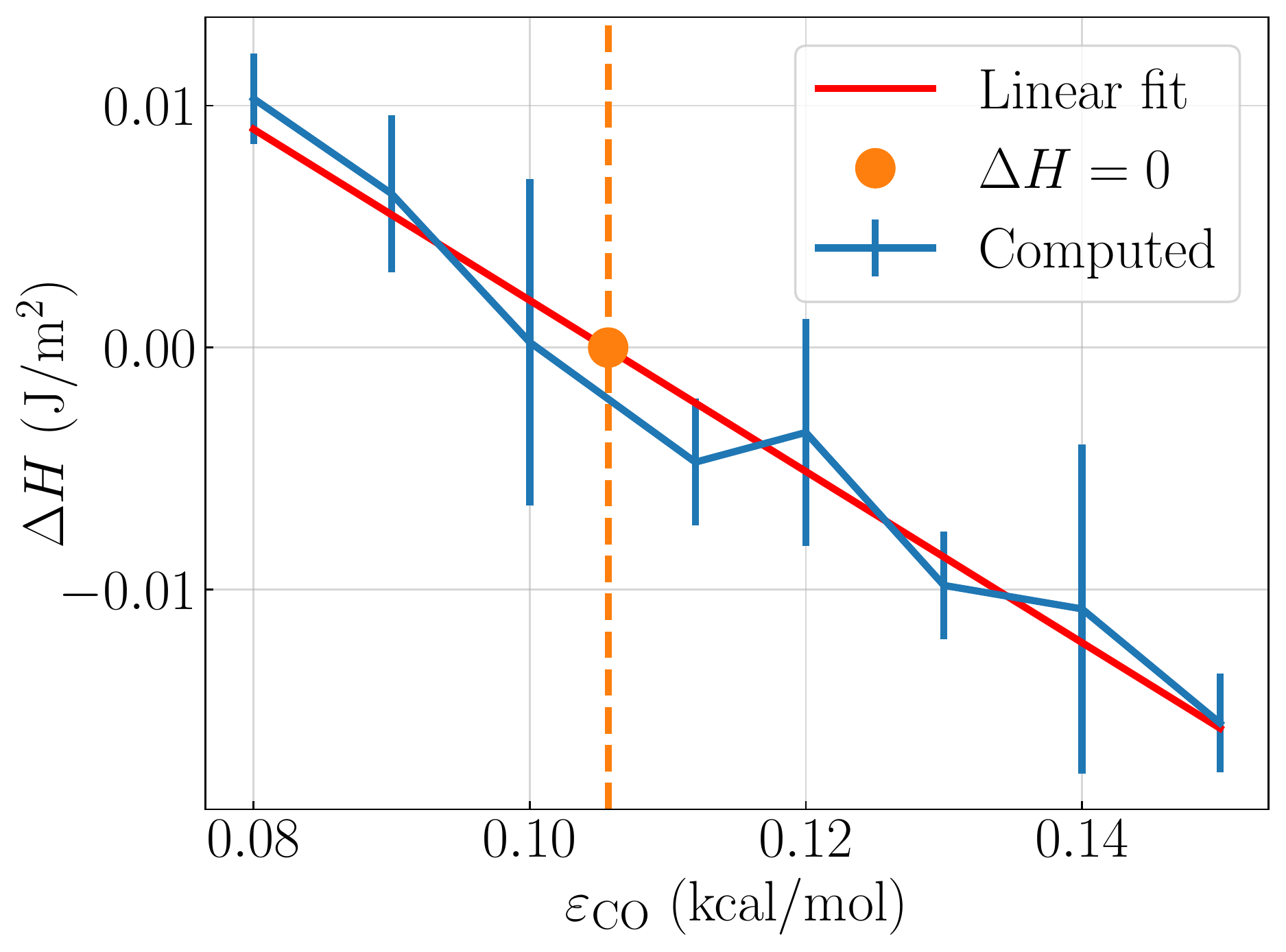}
    \caption{Enthalpy excess as a function of the carbon-oxygen interaction parameters $\varepsilon_\mathrm{CO}$ for a PG surface (blue line). The curve is fitted linearly (red curve) to determine the point at which $\Delta H=0$, which corresponds to $\varepsilon_\mathrm{CO}=0.105$ kcal/mol.}
    \label{fig:enthalpy_pristine}
\end{figure}
Another mean of computing the thermo-osmotic coefficient is to use Eq.~(\ref{eq:mto_nemd}), \textit{i.e.} generating a thermo-osmotic flow by applying a temperature gradient along the channel in the $x$ direction. Dealing with periodic boundary conditions in the $x$ direction, one can only apply the desired temperature gradient locally, not in the entire system. Here we divide the system in 2 regions: in region 1, we apply a temperature gradient $\nabla T_1$, and in region 2, we apply $\nabla T_2$, such that $L_1\nabla T_1=-L_2\nabla T_2=\Delta T$ (Fig.~\ref{fig:nemd_system}). Let us first consider an electrically neutral surface: the average velocity is given by the sum of a thermo-osmotic and a Poiseuille flow. It is uniform along the channel:
\begin{equation}
    \overline{v_\mathrm{tot}}=\overline{v_{\mathrm{to},1}}+\overline{v_{\mathrm{P},1}}=\overline{v_{\mathrm{to},2}}+\overline{v_{\mathrm{P},2}}
\end{equation}
where $\overline{v_{\mathrm{to},i}}$ is the average thermo-osmotic velocity over the channel thickness, and $\overline{v_{\mathrm{P},i}}$ is the average Poiseuille flow, in region $i$. The average Poiseuille velocity is given by \cite{herrero2022chapter}:
\begin{equation}\label{eq:avg_vpois}
    \overline{v_{\mathrm{P},i}}=-\frac{\nabla p_ih^2}{12\eta}\left(1+6\frac{b_i}{h}\right),
\end{equation}
where $h$ is the height of the channel, $b_i$ is the slip length, and $\nabla p_i$ is the pressure gradient in region $i$. $\nabla p_1=\Delta p/L_1$ and $\nabla p_2=-\Delta p/L_2$, with $\Delta p$ the pressure difference. For an electrically neutral surface, the thermo-osmotic flow rate is given by:
\begin{equation}
    \overline{v_{\mathrm{to},i}}=\frac{1}{h}\int_0^h v_{\mathrm{to},i}(z)\dd z.
\end{equation}
When the channel is large enough, \textit{i.e.} $h\gg\lambda_\mathrm{h}$, $v_\mathrm{to}(z)\approx v_\mathrm{to}(\infty)$, the average thermo-osmotic velocity becomes:
\begin{equation}
    \overline{v_{\mathrm{to},i}}=v_{\mathrm{to},\nabla T}^i=-M_{\mathrm{to},i}\frac{\nabla T_i}{T},
\end{equation}
where $\nabla T_i$ is the temperature gradient in region $i$; $\nabla T_1=\Delta T/L_1$ and $\nabla T_2=-\Delta T/L_2$, with $\Delta T$ the temperature difference. When the surfaces in region 1 and 2 are identical, conservation of flow rate predicts that Poiseuille and thermo-osmotic flows perfectly cancel each other, so that total flow is zero. One solution to this problem is to define the surface in region 2 such that $\Delta H=0$. We will call such a surface thermo-osmosis-insensitive (TOI). 

To obtain a TOI surface for the neutral PG wall, we start from a graphene surface and we slightly modify its solid-liquid interaction parameters, $\varepsilon_\mathrm{CO}\in[0.08, 0.15]$ kcal/mol in our case, which allows us to realize a surface with a null enthalpy excess for $\varepsilon_\mathrm{CO}=0.105$ kcal/mol (Fig.~\ref{fig:enthalpy_pristine}). Note that changing the $\varepsilon_\mathrm{CO}$ amounts to changing the contact angle (see Appendix~B for details on the calculation of the contact angle). 
We applied the same procedure for the polarized PG , neutral and charged GO surfaces: for each value of the surface charge or oxidation rate, we varied $\varepsilon_\mathrm{CO}$ to find the value canceling $\Delta H$. 

Considering a TOI surface, the equation of the average flow velocity becomes:
\begin{equation}
    \overline{v_\mathrm{tot}}=-M_\mathrm{to}\frac{\nabla T_1}{T}+\overline{v_{\mathrm{P},1}}=\overline{v_{\mathrm{P},2}},
\end{equation}
where we set $M_\mathrm{to,1}=M_\mathrm{to}$. 
\new{Using the expression of $\overline{v_{\mathrm{P},i}}$, Eq.~\eqref{eq:avg_vpois}, where $\nabla p_1=\Delta p/L_1$ and $\nabla p_2=-\Delta p/L_2$, one can obtain the expression of $\Delta p$: 
\begin{equation}
    \Delta p = \overline{v_\mathrm{tot}} \times \frac{12\eta L_2}{h^2 \left(1+6\,b_2/h\right)}, 
\end{equation}
and then relate the thermo-osmotic coefficient to the total flow velocity:}  
\begin{equation}\label{eq:mto_corrected}
    M_\mathrm{to}=-\frac{\overline{v_\mathrm{tot}}}{\nabla T_1/T}\left[1+\frac{L_2\left(1+6\,b_1/h\right)}{L_1\left(1+6\,b_2/h\right)}\right].
\end{equation}
When considering an electrically charged surface, one must take into account the permittivity-osmotic flow generated in region 2. The total flow rate becomes:
\begin{equation}
    \overline{v_\mathrm{tot}}=-M_\mathrm{to}\frac{\nabla T_1}{T}+\overline{v_{\mathrm{P},1}}=\overline{v_{\mathrm{to},\nabla\varepsilon,2}}+\overline{v_{\mathrm{P},2}},
\end{equation}
with
\begin{equation}
    \overline{v_{\mathrm{to},\nabla\varepsilon,2}}=v_{\mathrm{to},\nabla\varepsilon,2}^{\mathrm{PB}},
\end{equation}
if we consider the channel large enough compared to the EDL, $h\gg\mathrm{min}(\lambda_\mathrm{D},\ell_\mathrm{GC})$. The thermo-osmotic coefficient is then given by:
\begin{multline}
    M_\mathrm{to}=-\frac{1}{\nabla T_1/T}\Biggl\{(\overline{v_\mathrm{tot}}-v_{\mathrm{to},\nabla\varepsilon,2}^{\mathrm{PB}})\left[1+\frac{L_2\left(1+6\,b_1/h\right)}{L_1\left(1+6\,b_2/h\right)}\right]\\
    +v_{\mathrm{to},\nabla\varepsilon,2}^{\mathrm{PB}}\Biggr\}.
\end{multline}

We fixed $L_1\sim100$, $L_2\sim20$\,\AA{}, and $L_y\sim26$\,\AA{}. The size of the thermostated regions was set to 10\,\AA{}. We verified that the chosen temperature difference, $\Delta T=80$\,K, lied within the linear response regime.
\new{For NEMD simulations, we performed only one measure per data point; the uncertainty originates in the uncertainty on the enthalpy excess of the thermo-osmosis-insensitive surface.}

\section{Results and discussion}\label{sec:results}

\begin{figure}
    \centering
    \includegraphics[width=0.45\textwidth]{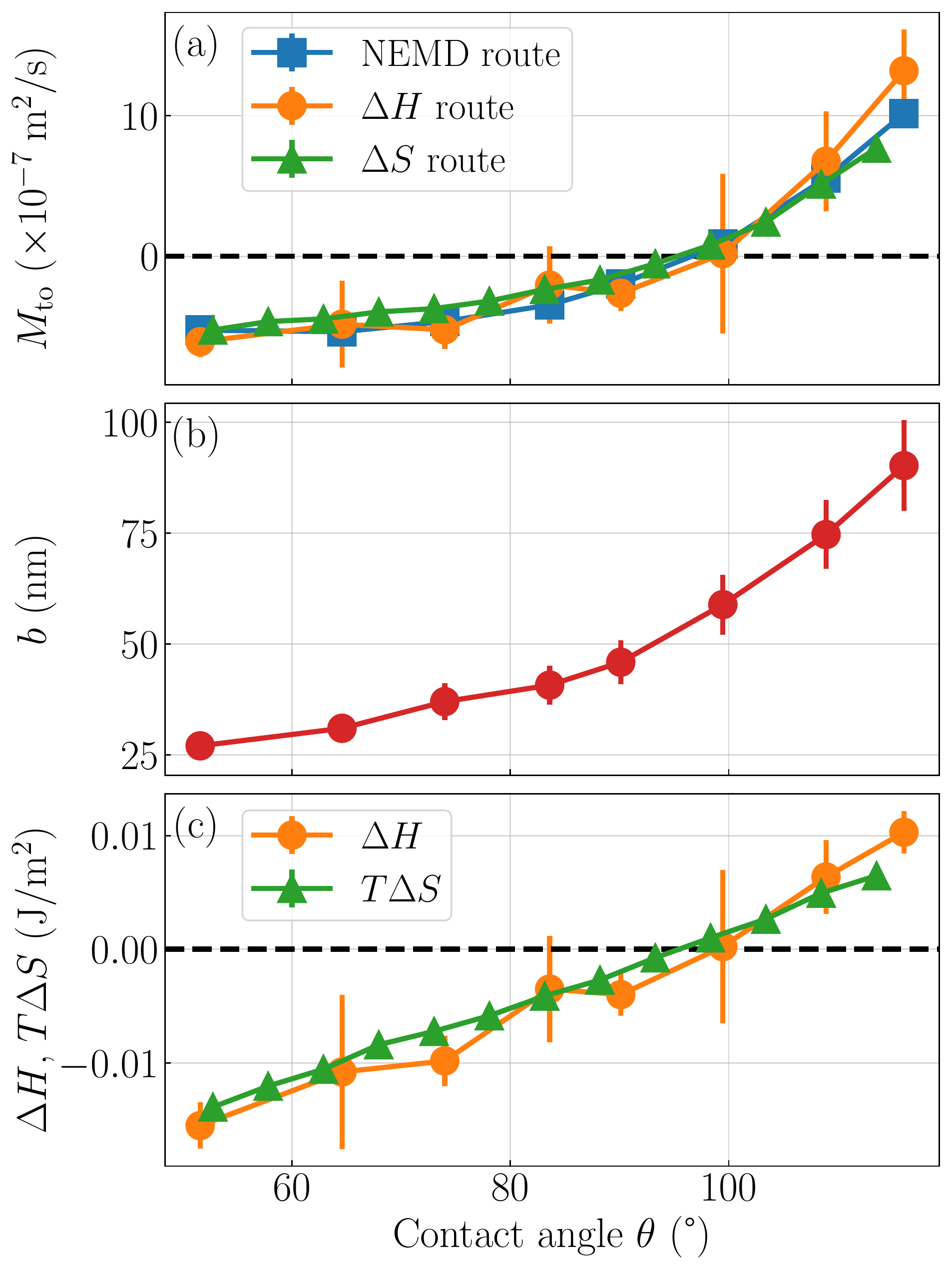}
    \caption{Thermo-osmotic coefficient of a pristine graphene-like surface with varying wetting properties, as a function of the contact angle. In figure a), the blue squares represent the values of the NEMD simulations, wherein a flow is generated by applying a thermal gradient along the interface (Eq.~\ref{eq:mto_corrected}). The orange circles represent the curve generated by the computation of the enthalpy excess and the slip length, following Derjaguin's approach (Eq.~\ref{eq:mto_derjaguin}). The green triangles represent the results of the computation of the entropy excess through the dry-surface method (DSM) (Eq.~\ref{eq:mto_entropy}). Panel b) represents the slip length. Panel c) represents the enthalpy excess (orange circles) and the entropy excess (green triangles). \new{Some error bars do not appear because they are smaller than the symbols.}}
    \label{fig:pg_neutral}
\end{figure}

\begin{figure}
    \centering
    \includegraphics[width=0.45\textwidth]{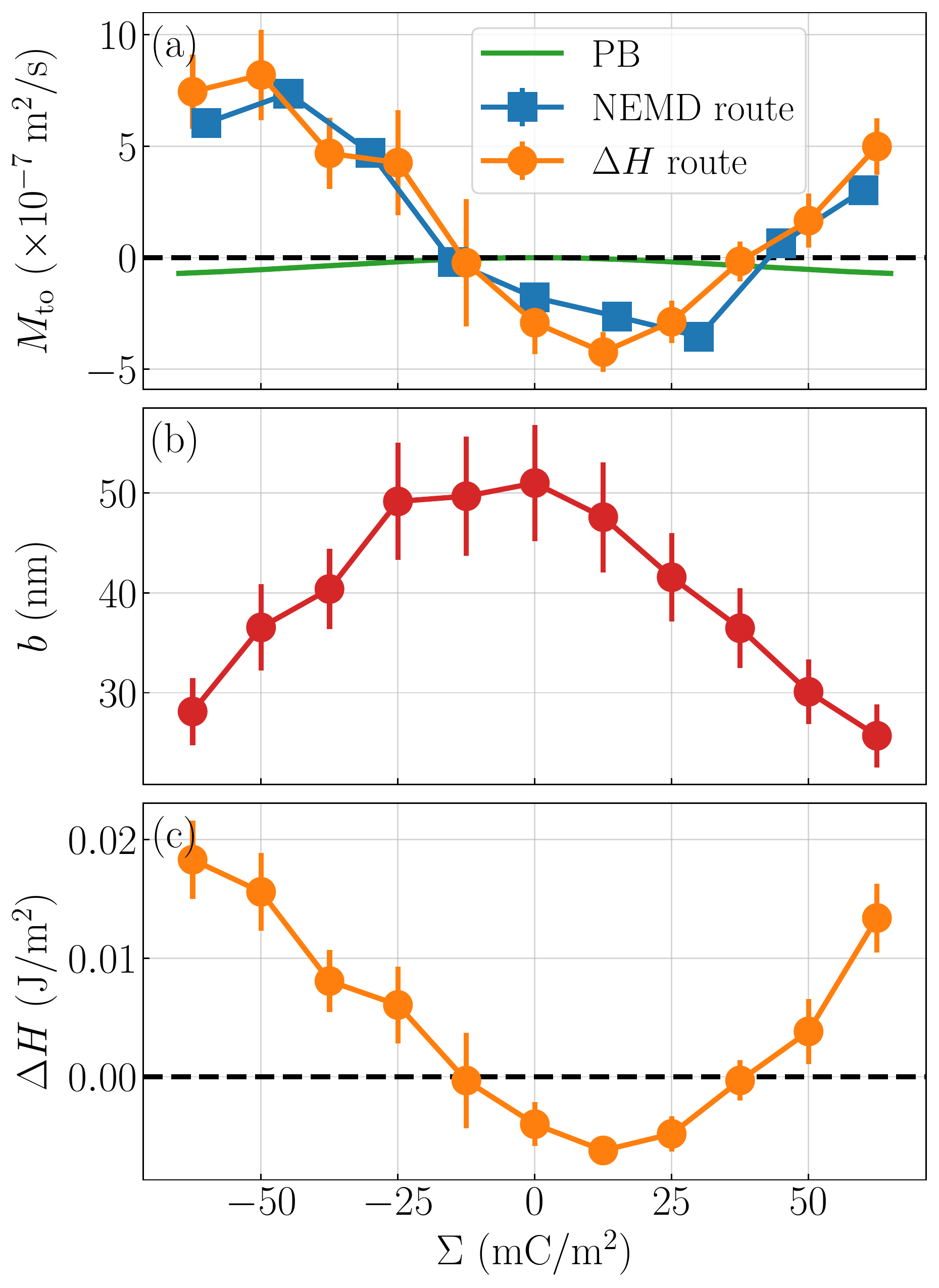}
    \caption{Thermo-osmotic coefficient of a polarized graphene surface as a function of the surface charge density. The non-equilibrium computation (blue squares) agrees well with Derjaguin's formula based on the enthalpy excess (orange circles). The Poisson-Boltzmann (PB) prediction to thermo-osmosis is also represented (green line). As discussed in the text, it is negligible compared to the contribution of water enthalpy excess. Panels b) and c) represent respectively the slip length and the enthalpy excess. \new{Some error bars do not appear because they are smaller than the symbols.}}
    \label{fig:pg_charged}
\end{figure}

\begin{figure}
    \centering
    \includegraphics[width=0.45\textwidth]{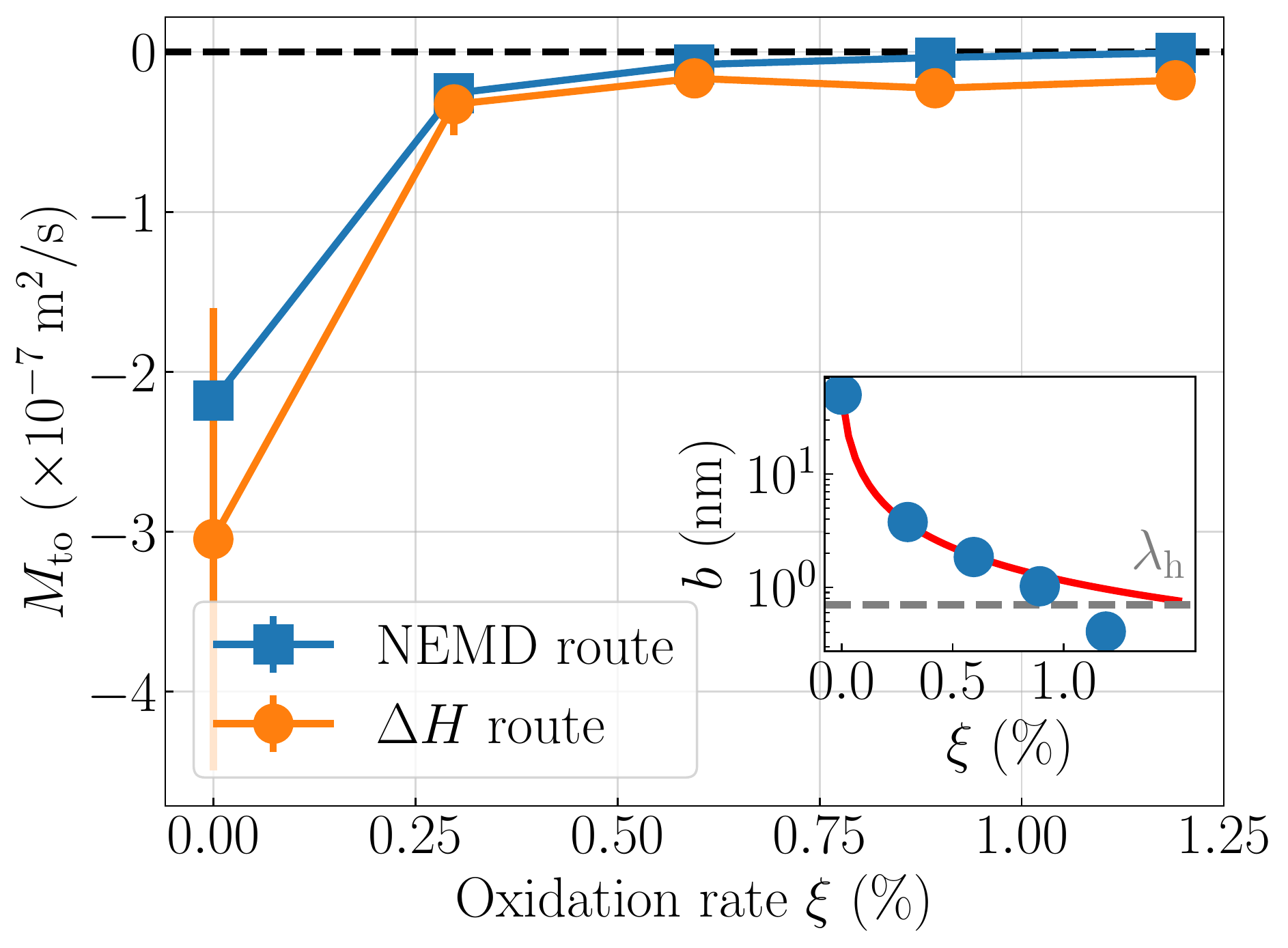}
    \caption{Thermo-osmotic coefficient of a neutral graphene oxide surface as a function of the oxidation rate. We used two methods to compute the thermo-osmotic coefficient, non-equilibrium simulations (blue squares) and the enthalpy excess route (orange circles). The inset shows the slip length as a function of the oxidation rate (blue circles), the red curve represents the prediction of Eq.~(\ref{eq:slip_go_neutral}) with $\sigma_\mathrm{h}=1.2$\,\AA{}. The thickness of the interaction layer $\lambda_\mathrm{h}$ is shown with a dotted line for comparison. \new{Some error bars do not appear because they are smaller than the symbols.}}
    \label{fig:go_neutral}
\end{figure}

\begin{figure}
    \centering
    \includegraphics[width=0.45\textwidth]{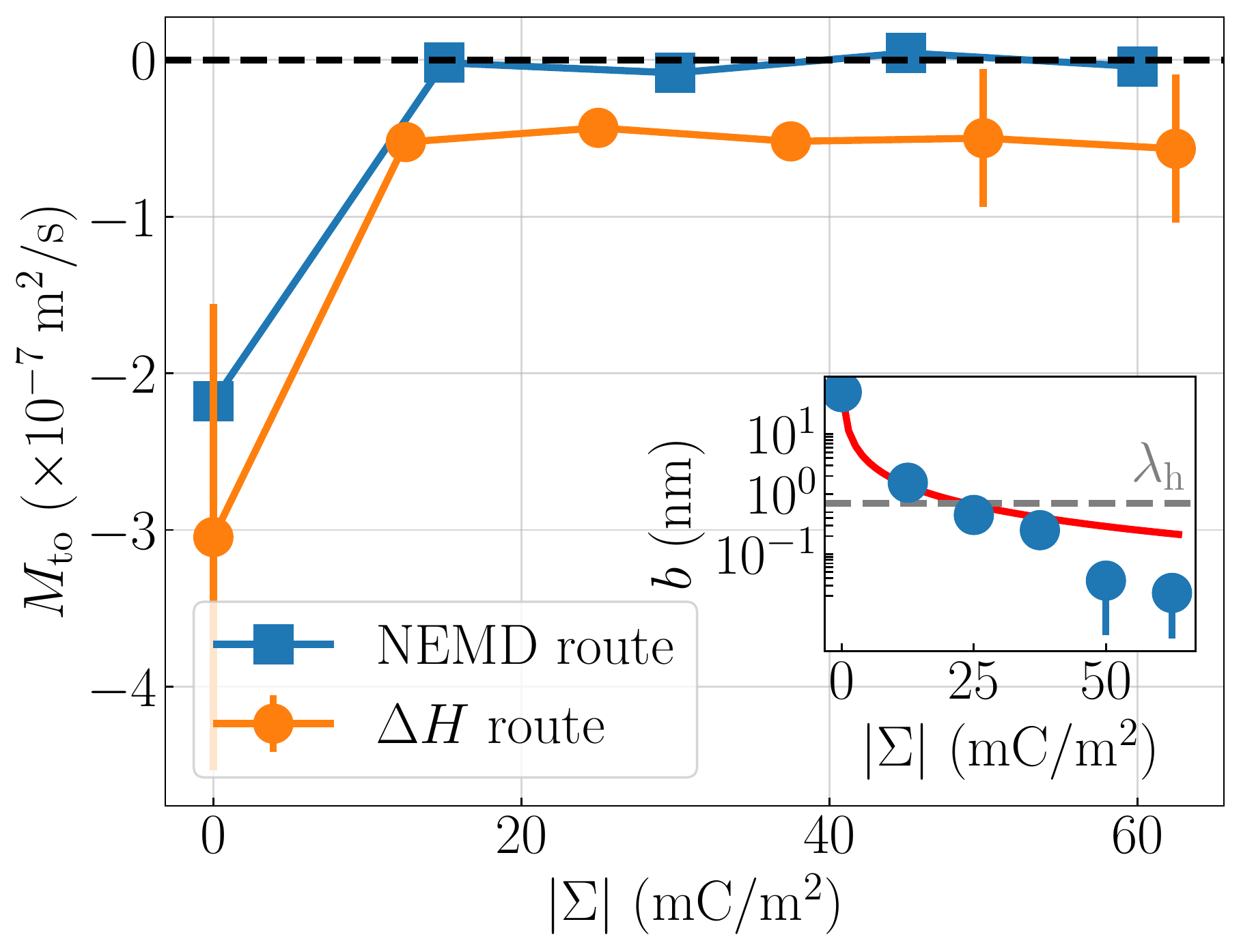}
    \caption{Thermo-osmotic coefficient as a function of the surface charge density for a charged graphene oxide surface. The blue squares represent the non-equilibrium calculation and the orange circles shows the enthalpy excess route. Inset: slip length as a function of the surface charge density in blue circles, theoretical prediction in red (using Eq.~(\ref{eq:slip_go_charged}) with $\sigma_\mathrm{h}=8.8$\,\AA{}). The thickness of the interaction layer $\lambda_\mathrm{h}$ is shown with a dotted line for comparison. \new{Some error bars do not appear because they are smaller than the symbols.}}
    \label{fig:go_charged}
\end{figure}

\begin{figure}
    \centering
    \includegraphics[width=0.5\textwidth]{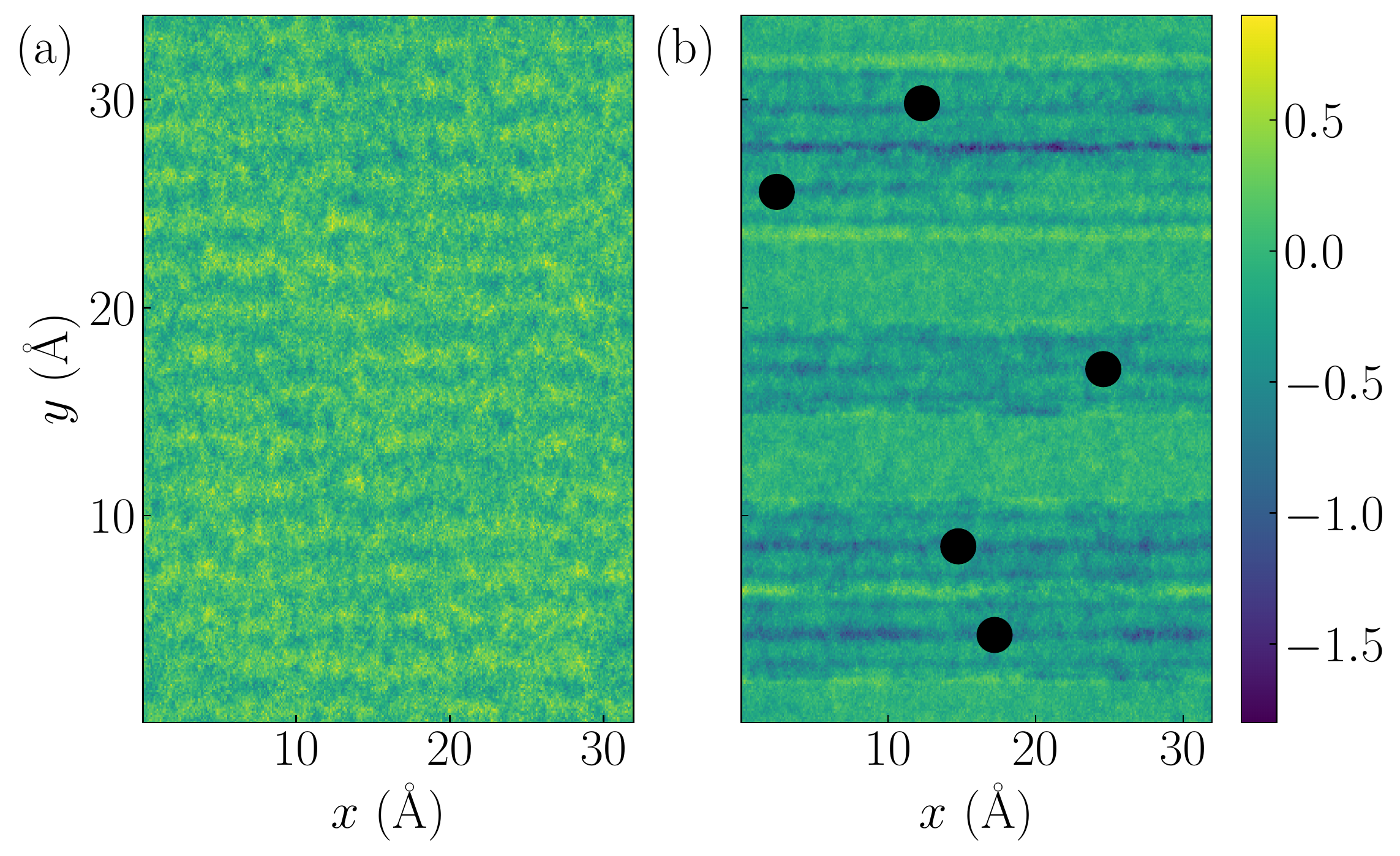}
    \caption{Enthalpy excess map for a pristine graphene surface (a) and for a graphene oxide surface (b), the black points represent the position of the hydroxyl groups. Units of the colorbar are in J/m$^2$.}
    \label{fig:map_enthalpy}
\end{figure}

\subsection{Comparison of the different approaches}
We begin by comparing the different methods used to compute the thermo-osmotic coefficient and we assess their agreement. Figure~\ref{fig:pg_neutral} presents the thermo-osmotic coefficient $M_\mathrm{to}$, on a graphene-like surface with tunable wetting properties, as a function of the contact angle. We found a remarkable agreement between Derjaguin's approach (both through the enthalpy excess route and the entropy excess route) and the direct computation of the thermo-osmotic coefficient using NEMD. The calculations based on the enthalpy or the entropy excess give additional information on the contributions to $M_\mathrm{to}$, \textit{i.e.} the driving force $\nabla\gamma=-\Delta H\nabla T/T=-\Delta S\nabla T$, and hydrodynamic properties (the slip length and viscosity). The agreement between the calculation of entropy excess with the DSM and the computation of enthalpy excess, see Fig.~\ref{fig:pg_neutral}.c, tends to validate the computation of the enthalpy excess using the pressure tensor. The DSM is much more accurate than the direct enthalpy excess calculation, while being faster in terms of computational time. A first limitation of the DSM to compute the entropy excess is the presence of a parameter in Eq.~(\ref{eq:dS_dsm}), the solid-liquid entropy of the reference state $\kappa_0$. We determined this value by finding the point at which the entropy excess is zero, for which $M_\mathrm{to}$ vanishes (based on the excess enthalpy route results). Another limitation of this method is that it cannot be used with electrically charged surface. Indeed, it requires turning off the solid-liquid Coulombic interactions, which in practice means removing the charge of the wall, leading to non-neutral systems. In contrast, both NEMD and $\Delta H$ route can be used on charged surfaces, and provide consistent results (Fig.~\ref{fig:pg_charged}).

Returning to the NEMD computation, the agreement between this method and the methods based on the computation of the enthalpy and the entropy excess tends to confirm the validity of Derjaguin's approach to compute the thermo-osmotic coefficient for smooth surfaces.
\new{In particular, Eq.~\eqref{eq:osmotic_velocity} assumes that the viscosity is homogeneous, and deviations from this assumption should have a strong impact on the osmotic response, see Appendix~C. Therefore, our results tend to show the viscosity does not vary significantly close to the walls considered here. This is compatible with previous work showing a constant viscosity close to hydrophobic walls \cite{bonthuis_beyond_2013}.}
NEMD is comparable to enthalpy excess computation in terms of computational time while being much more precise. On the other hand, it requires finding a TOI surface, \textit{i.e.}, computing the enthalpy excess for a range of interaction parameters and finding the surface that nullifies the enthalpy excess. Another limitation of the method is that it provides limited microscopic information, as it only returns one quantity, $M_\mathrm{to}$.

The agreement between the methods remains consistent for the neutral GO surface (Fig.~\ref{fig:go_neutral}). However, for the charged GO surface, there is a clear departure from Derjaguin's theory (Fig.~\ref{fig:go_charged}). Two reasons could explain this difference; the first reason is the assumption that viscosity is constant along the channel. While this assumption is reasonable for smooth surfaces, it may not be true for defective ones. In Ref.~\citenum{sendner_interfacial_2009}, the authors computed the velocity profile of water near diamond interfaces by generating a Couette flow in the channel. They considered two types of surfaces: a smooth diamond surface and a defective surface with hydroxyl groups as defects. They found that the presence of hydroxyl groups could significantly increase the viscosity near the surface. Consequently, the thermo-osmotic velocity predicted by Derjaguin's formula could be overestimated\new{, see Appendix C and Ref.~\citenum{ouadfel_complex_2023}}.

Another possible reason involves 
the distribution of enthalpy excess. In Fig.~\ref{fig:map_enthalpy}, we computed the map of enthalpy excess for both the pristine and defective surfaces \new{using a NEMD simulation with a flow along the $x$ direction}. While the enthalpy excess of the smooth surface is homogeneous on the $xy$ plane, it becomes heterogeneous with the GO surface. Specifically, the enthalpy excess is not constant along the $y$ direction. The non-uniform enthalpy excess near the surface could explain the deviation from Derjaguin's theory, which assumes a uniform enthalpy excess.

\subsection{Thermo-osmosis on homogeneous and heterogeneous surfaces}
Now, let us examine the physical interpretation of the results on thermo-osmosis on graphene-like surfaces. In Fig.~\ref{fig:pg_neutral}, one can observe that the thermo-osmotic coefficient varies considerably with the different wettability conditions considered. For hydrophilic surfaces, the flow is thermophilic, \textit{i.e.} the liquid flows toward the hot side. However, for hydrophobic surfaces, the flow is thermophobic. The change of sign in the thermo-osmotic coefficient when modifying the wetting has been observed experimentally \cite{bregulla_thermo-osmotic_2016} and predicted with MD simulations \cite{fu_what_2017,ouadfel_complex_2023} in other systems. The change of direction of the thermo-osmotic flow can be understood by the change of sign of the enthalpy excess (Fig.~\ref{fig:pg_neutral}.c).

The thermo-osmotic coefficient computed here is one order of magnitude greater than that of previous MD works: it was found to be around $10^{-8}$\,m$^2$/s for model LJ systems \cite{fu_what_2017} and water-silica nanochannel \cite{chen_thermo-osmosis_2023}. Although the enthalpy excess values are similar across the systems, the exceptionally high slip length of water on graphene surfaces leads to a massive amplification of the thermo-osmotic flow.

In Fig.~\ref{fig:pg_charged}, one can see a significant impact of the surface charge density on the thermo-osmotic coefficient, with a fluid flow enhancement by a factor of $\sim$ 3 for the highest value of $M_\mathrm{to}$ compared to the electrically neutral graphene surface. Once again, we observe a change of sign when changing the surface charge density, and this behavior can be fully explained by the variation of the enthalpy excess with $\Sigma$. The variation of the enthalpy excess has already been characterized in Ref.~\citenum{ouadfel_complex_2023} for a model system, it is parabolic-like. For the pristine graphene considered here, the enthalpy excess is negative for $\Sigma=0$ mC/m$^2$, which gives two points where $\Delta H=0$, and thus two points for which the thermo-osmotic flow changes its direction.
Finally, as stated in Sec.~\ref{sec:theory}, the Poisson-Boltzmann (PB) contribution to the enthalpy excess, and thus the thermo-osmotic coefficient, is negligible compared to the contribution of water molecules.

Figures \ref{fig:go_neutral} and \ref{fig:go_charged} represent the thermo-osmotic coefficient for a neutral and a charged GO surface, respectively, along with their slip length. When considering defective surfaces, the presence of defects greatly reduces the solid-liquid slip \cite{mangaud_chemisorbed_2022}, thereby diminishing the osmotic flow generated. The neutral GO surface still exhibits a relatively high thermo-osmotic coefficient for very small oxidation rates. The attenuation is more pronounced for charged defects, as the electrostatic interactions increase solid-liquid friction. In this case, the fluid flow quickly drops to values that are too small to be measured with our method.

The variation of the slip length with the surface density is given in Ref.~\citenum{xie_liquid-solid_2020} for a heterogeneous surface:
\begin{equation}\label{eq:slip_go_charged}
    b=\frac{b_0}{1+3\pi\sigma_\mathrm{h}b_0|\Sigma|/q},
\end{equation}
with $b_0$ the slip length of the neutral surface, $\sigma_\mathrm{h}$ the hydrodynamic radius of the defect plus counter-ion assembly, treated here as a fitting parameter. We used $\sigma_\mathrm{h}=8.8$\,\AA{} for our purpose. Similarly, for a neutral defective surface, one can write the slip length as:
\begin{equation}\label{eq:slip_go_neutral}
    b=\frac{b_0}{1+3\pi\sigma_\mathrm{h}b_0n_\mathrm{C}\xi},
\end{equation}
with $n_\mathrm{C}$ the number density of the carbon atoms, and $\sigma_\mathrm{h}=1.2$\,\AA{} the hydrodynamic radius of the hydroxyl groups. In both cases, the slip length quickly decreases even at low oxidation degree. Therefore, one can expect the experimental measured thermo-osmotic coefficient of a graphene nanochannel to be lower than what we have found numerically, due to the presence of defects.

\section{Conclusion}
In this paper, we have presented three methods to compute the thermo-osmotic coefficient using molecular simulations. The first two are quite similar in that they are based on Derjaguin's theoretical framework. The first one requires to compute the enthalpy excess at equilibrium. The second one uses the dry-surface method to compute the solid-liquid entropy, which in linked to the enthalpy excess. The last method is based on NEMD, the thermo-osmotic velocity is computed directly by applying a thermal gradient in the liquid along the solid-liquid interface. We found a good agreement between the three methods, validating the relevance of Derjaguin's approach to study thermo-osmosis, for homogeneous surfaces. Even though the methods agree one with each other, they require different simulation times and are not equally accurate. For instance, the computation of the solid-liquid entropy using dry-surface method is much more accurate than the calculation of the enthalpy excess, requires less computational time, but is limited to electrically neutral surfaces. The NEMD calculation is very precise but requires to find a surface with a vanishing enthalpy excess, increasing the computational cost of the method.

We applied this methods to study the thermo-osmotic slip of charged graphitic surface. Although the neutral pristine graphene surface displays a relatively large thermo-osmotic coefficient due to its high slip length, we found that polarizing the graphene surface could significantly enhance the strength of the thermo-osmotic flow, by a factor 3 to 4 for the highest values. We found that this amplification is due to the variation of the enthalpy excess with the surface charge density. Our findings suggest that by controlling the surface charge density, one can precisely make the liquid flow in either directions in the channel, or not flow at all. Overall, these results call for experimental exploration of thermo-osmotic flow in the vicinity of electrically charged graphene surfaces.

\begin{acknowledgments}
The authors thank D. Pandey and S. Hardt for fruitful discussions. 
This work was supported by the ANR, Project ANR-21-CE50-0042 smoothE. 
This work used the HPC resources from the CNRS/IN2P3 Computing Center (Lyon - France) and from the PSMN mesocenter in Lyon. This manuscript is distributed under a Creative Commons Attribution $|$ 4.0 International license. 
\end{acknowledgments}

\section*{Data availability}

The data that support the findings of this study are available from the corresponding authors upon reasonable request.

\section*{Appendix A: Equivalence between entropy excess and solid-liquid entropy}\label{sec:appendixA}
In the framework of surface thermodynamics \cite{rowlinson_molecular_2002}, the solid-liquid entropy can be written as:
\begin{equation}
    s_\mathrm{sl}=\int_0^\infty\dd z\left[\sum_i(n_i(z)s_i(z)-n_i^\mathrm{bulk}s_i^\mathrm{bulk})\right].
\end{equation}
The entropy excess $\Delta S$ is then expressed as:
\begin{align}
    \Delta S &= s_\mathrm{sl}-\sum_i s_i^\mathrm{bulk}\int_0^\infty\dd z(n_i(z)-n_i^\mathrm{bulk})\\
    &= s_\mathrm{sl}-\sum_i s_i^\mathrm{bulk}\Gamma_i,
\end{align}
where $\Gamma_i$ is the adsorption of species $i$. In principle, there is a difference between $\Delta S$ and $s_\mathrm{sl}$, and this difference depends on the wetting properties of the surface, as the adsorption depends on the wetting. However, in practice, the calculation of the solid-liquid entropy matches very well with the enthalpy excess route, and with NEMD results (Fig.~\ref{fig:pg_neutral}), suggesting an equivalence between $\Delta S$ and $s_\mathrm{sl}$.

\section*{Appendix B: Contact angle computation}\label{sec:appendixB}
To characterize the wetting of the solid-liquid interface, we performed simulations to compute the contact angle of liquid droplets on the graphene-like surfaces. We consider two systems of different size to quantify finite-size effects. The first system featured 1000 water molecules enclosed between graphene-like walls of size $L_x\sim L_y\sim100$\,\AA{}. The walls were separated by a distance of 60\,\AA{}, to ensure that the droplet formed on one wall does not interact with the other wall. The second system was composed of 4000 water molecules. The size of walls was $L_x\sim L_y\sim180$\,\AA{}. They were kept at a distance of 100\,\AA{}. We ran an equilibration phase for 0.5\,ns, the production period lasted 4\,ns. We computed the contact angle by fitting the average density profile along $z$ by \cite{fu_giant_2019-1}:
\begin{equation}
    \rho_\mathrm{app}(z)=\frac{(\rho_\mathrm{l}-\rho_\mathrm{v})\pi r(z)^2}{L_xL_y}+\rho_\mathrm{v},
\end{equation}
where $\rho_\mathrm{app}$ is the apparent density profile, $\rho_\mathrm{l}$ and $\rho_\mathrm{v}$ are the real liquid and vapor densities, and $r(z)$ is the radius of the droplet section, which is equal to $\sqrt{R^2-(z-z_0)^2}$, with $R$ the radius and $z_0$ the center of the sphere. The contact angle is then given by:
\begin{equation}
    \theta=\mathrm{arccos}\left(\frac{z_\mathrm{s}-z_0}{R}\right),
\end{equation}
where $z_\mathrm{s}$ is the hydrodynamic wall position \cite{herrero_shear_2019}.

\begin{figure}
    \centering
    \includegraphics[width=0.45\textwidth]{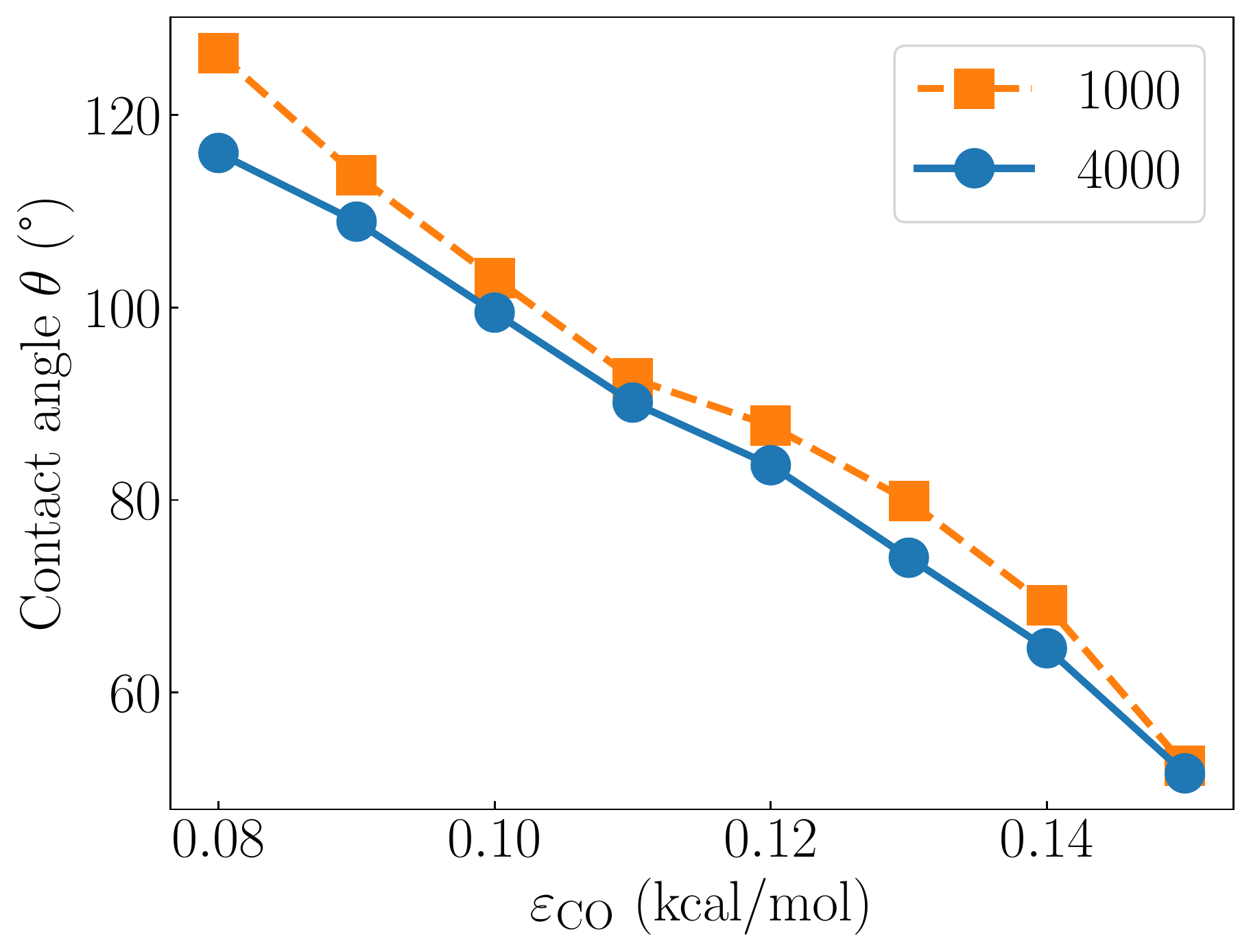}
    \caption{Contact angle as a function of the liquid-wall interaction parameter. We compare the contact angle for two size of droplets, 1000 water molecules (orange squares) and 4000 water molecules (blue circles).}
    \label{fig:contact_angle}
\end{figure}

The variation of the contact angle with $\varepsilon_\mathrm{CO}$ is presented in Fig.~\ref{fig:contact_angle}. The contact angle values given in the main text are for the 4000-molecule water droplet, which are more accurate because they are less affected by finite-size effects.

\section*{Appendix C: Osmotic velocity with a heterogeneous viscosity}\label{sec:appendixC}

\new{In this appendix, we provide extensions of Eq.~\eqref{eq:osmotic_velocity}, which relates the osmotic velocity to the force density exerted on the fluid in the $x$ direction, for a heterogeneous (local) viscosity. 
Denoting $z$ the distance to the wall, the Stokes equation writes: 
\begin{equation}\label{eq:stokes_etaz}
    -\eta(z) \frac{\dd^2 v}{\dd z^2} = f_x(z), 
\end{equation}
with $\eta(z)$ the viscosity, $f_x(z)$ the external force density, and $v(z)$ the velocity along the $x$ direction.} 

\new{Denoting $b_0$ the local slip length at the wall, $b_0 = \eta(0)/\lambda$, with $\lambda$ the interfacial friction coefficient, the slip boundary condition at the wall writes: 
\begin{equation}
    v(0) = b_0 \left. \frac{\dd v}{\dd z} \right|_{z=0} . 
\end{equation}
Denoting $\eta_\text{bulk}$ the viscosity far from the wall, one can solve Eq.~\eqref{eq:stokes_etaz} with the slip boundary condition and a vanishing shear rate far from the wall to obtain the osmotic velocity, i.e. the plateau of velocity far from the wall: 
\begin{equation}
        v_\mathrm{osm} = \frac{1}{\eta_\text{bulk}} \int_0^\infty (z+b_0) f_x(z) \frac{\eta_\text{bulk}}{\eta(z)} \dd z .
\end{equation}}

\new{Similarly, for a no-slip surface with a stagnant layer of thickness $z_\text{s}$, the osmotic velocity writes: 
\begin{equation}
        v_\mathrm{osm} = \frac{1}{\eta_\text{bulk}} \int_{z_\text{s}}^\infty (z-z_\text{s}) f_x(z) \frac{\eta_\text{bulk}}{\eta(z)} \dd z .
\end{equation}}

\new{Therefore, any local increase in the viscosity amounts to effectively reducing the driving force density by a factor $\eta(z)/\eta_\text{bulk}$ as compared to Eq.~\eqref{eq:osmotic_velocity} assuming a homogeneous viscosity.}

\section{References}

%


\end{document}